\documentclass[12pt]{article}
\pdfoutput=1

\usepackage{jheppub}
\usepackage{amsmath,amssymb}
\usepackage{caption}
\usepackage{subcaption}
\usepackage[per-mode=symbol]{siunitx}
\usepackage{bm}
\usepackage{soul}
\usepackage{todonotes}
\usepackage[toc,page]{appendix}
\usepackage{tocloft}
\usepackage[autostyle=true]{csquotes}

\usepackage{parskip}
\newtheorem{definition}{Definition}[section]
\newtheorem{theorem}{Theorem}[section]

\newtheorem{conjecture}[theorem]{Conjecture} 

\newtheorem{example}{Example}

\newcommand\be{\begin{equation}}
\newcommand\ee{\end{equation}}
\newcommand\bea{\begin{eqnarray}}
\newcommand\eea{\end{eqnarray}}
\newcommand{\cX}{{\mathcal X}}
\renewcommand\tilde{\widetilde}

\title{\begin{center}New Calabi--Yau Manifolds\\ from Genetic Algorithms\end{center}}
	
\author[a]{Per Berglund,}
\author[b,c,d,e]{Yang-Hui He,}
\author[b,c]{Elli Heyes,}
\author[f]{Edward Hirst,}
\author[g]{Vishnu Jejjala,}
\author[h]{Andre Lukas}

\affiliation[a]{Department of Physics and Astronomy, University of New Hampshire, Durham, NH 03824, USA}
\affiliation[b]{London Institute for Mathematical Sciences, Royal Institution, London, W1S 4BS, UK}
\affiliation[c]{Department of Mathematics, City, University of London, EC1V 0HB, UK}
\affiliation[d]{Merton College, University of Oxford, OX1 4JD, UK}
\affiliation[e]{School of Physics, NanKai University, Tianjin, 300071, P.R.\ China}
\affiliation[f]{Centre for Theoretical Physics, 
Queen Mary University of London, E1 4NS, UK}
\affiliation[g]{Mandelstam Institute for Theoretical Physics, School of Physics, NITheCS, and CoE-MaSS, University of the Witwatersrand, Johannesburg, WITS 2050, South Africa}
\affiliation[h]{Rudolf Peierls Centre for Theoretical Physics, University of Oxford, OX1 3PU, UK\\}

\emailAdd{per.berglund@unh.edu}
\emailAdd{hey@maths.ox.ac.uk}
\emailAdd{elli.heyes@city.ac.uk}
\emailAdd{e.hirst@qmul.ac.uk}
\emailAdd{v.jejjala@wits.ac.za}
\emailAdd{andre.lukas@physics.ox.ac.uk}

\preprint{\begin{flushright}
QMUL-PH-23-06
\end{flushright}}

\abstract{
Calabi--Yau manifolds can be obtained as hypersurfaces in toric varieties built from reflexive polytopes. We generate reflexive polytopes in various dimensions using a genetic algorithm. As a proof of principle, we demonstrate that our algorithm reproduces the full set of reflexive polytopes in two and three dimensions, and in four dimensions with a small number of vertices and points. Motivated by this result, we construct five-dimensional reflexive polytopes with the lowest number of vertices and points.
By calculating the normal form of the polytopes, we establish that many of these are not in existing datasets and therefore give rise to new Calabi--Yau four-folds. In some instances, the Hodge numbers we compute are new as well.
}

\begin{document}
\maketitle

\section{Introduction}\label{sec:intro}
Ever since the early days of string phenomenology~\cite{Candelas:1985en}, the construction of Calabi--Yau (CY) manifolds had been an active pursuit in the theoretical physics and mathematics communities.
(See, for example, the classic reference~\cite{Hubsch:1992nu} as well as the recent textbooks~\cite{He:2018jtw,Bao:2020sqg}.)
This is motivated by the existence of Ricci-flat metrics on CY manifolds, a property which makes them a useful building block for compactifications of string theory.
In particular, CY three-folds compactify ten-dimensional superstring theory to four-dimensional quantum field theories with $\mathcal{N}=1$ supersymmetry~\cite{Candelas:1985en}.

In a parallel, and initially seemingly unrelated, vein,
lattice polytopes play a central role in geometry, where a polytope defines a fan of strongly convex rational polyhedra cones which in turn defines a toric variety.
In~\cite{batyrev1993dual,batyrev1994calabiyau}, Batyrev and Borisov showed how mirror pairs of $(n-1)$ complex dimensional CY manifolds can be realized from hypersurfaces in toric varieties constructed from $n$-dimensional reflexive polytopes.

Threading these two directions of CY compactifications and toric hypersurfaces, and motivated by their utility for string phenomenology, Kreuzer and Skarke devised an algorithm to generate all reflexive polytopes in $n$ dimensions~\cite{Kreuzer_1997}.
The algorithm consists of two steps.
First, a set $S$ of ``maximal'' polytopes is constructed such that any reflexive polytope is a subpolytope of a polytope in $S$.
These maximal polytopes are defined by a so-called ``weight system'' or a combination of weight systems.
Second, all subpolyhedra of all polyhedra in $S$ are constructed and checked for reflexivity.
The complete classification of $4319$ three-dimensional reflexive polytopes with this algorithm was accomplished in~\cite{kreuzer1998classification}.
From these we obtain K3 surfaces, which is to say, CY two-folds.
Proceeding to dimension three,
the $184,026$ weight systems giving rise to four-dimensional reflexive polytopes were presented in~\cite{skarke_1996}, and the resulting $473,800,776$ four-dimensional reflexive polytopes, leading to CY three-folds,  were listed in~\cite{kreuzer2000complete}.
In five dimensions the total number of reflexive polytopes is prohibitively large, and Sch\"oller and Skarke were only able to run the first stage of the algorithm to calculate all $322,383,760,930$ weight systems corresponding to maximal polytopes~\cite{Sch_ller_2019}.
They found that $185,269,499,015$ of these weight systems give rise to reflexive polytopes directly.\footnote{It can be that distinct weight systems correspond to the same polytope. This means that the total number of unique five-dimensional reflexive polytopes found by Sch\"oller and Skarke is less than $185,269,499,015$ (it is not known by how much).}
This result constitutes a partial classification, and we will compare our results to this list later on.\footnote{All data produced by the Kreuzer--Skarke algorithm for reflexive polytopes in three and four dimensions can be found at~\cite{KSweb}. The complete Sch\"oller-Skarke dataset of five-dimensional reflexive polytopes can be found at~\cite{SSweb}.}
The CY four-folds obtained from five-dimensional reflexive polytopes facilitate F-theory model building.

Finding reflexive polytopes is not an easy task.
A lattice polytope is reflexive when it satisfies a set of conditions:
it must have only a single interior point (the so-called IP property), its dual must as well be a lattice polytope (that is, its vertices must lie on integer lattice points), and the dual must also satisfy the IP property.
Alternatively and equivalently, a polytope is reflexive if and only if it satisfies the IP property and all bounding hyperplanes of the polytope lie at unit hyperplane distance from the origin. 
With multiple criteria, regression is unlikely to perform well at finding reflexive polytopes.
Any loss function will have local minima, where polytopes satisfy some but not all of the conditions for reflexivity.
We can ask if methods such as reinforcement learning or genetic algorithms, which explore an ``environment" in order to maximize a fitness or reward function, might be better suited for such a task.
In this paper, we will address this question for genetic algorithms (GAs) and leave reinforcement learning for future work.
Significant work has already been done on applying machine learning techniques to study objects in string theory, including polytopes~\cite{bao2021polytopes,Berman:2021mcw,Berglund:2021ztg}.
Genetic algorithms in particular have been successful at scanning for phenomenologically attractive string models~\cite{Abel_2014,abel2021string} and cosmic inflation models~\cite{Abel_2022}, but this is the first time that genetic algorithms have been used to search for reflexive polytopes.

We note that alongside classifying reflexive polytopes, there has also been focus on classifying $n$-dimensional smooth Fano polytopes, which correspond to reflexive polytopes with the extra condition that there are exactly $n$ edges emanating from each vertex and the primitive edge directions form a lattice basis of $\mathbb{Z}^{n}$. The classification of smooth Fano polytopes, up to dimension 9, was carried out~\cite{MR3792741} using the polymake software~\cite{MR2721537}. Using this database there was some recent work done on generating smooth Fano polytopes~\cite{nodland2022} using sequential modeling.
However, applying this methodology to reflexive polytopes without the smoothness condition does not yield good results.

The organization of this paper is as follows.
In Sections~\ref{sec:refpol} and~\ref{sec:GA}, we present some background on reflexive polytopes and briefly review genetic algorithms.
In Section~\ref{sec:lower_dim_results} we present the results of our GA searches for reflexive polytopes in two, three, and four dimensions and compare the results to the known complete classifications.
Five-dimensional reflexive polytopes are tackled in  Section~\ref{sec:5d_results}.
Using GAs, we generate datasets of five-dimensional reflexive polytopes with the smallest number of points and vertices.
From these we extract the polytopes with $h^{1,1}=1$ and $h^{1,1}=2$ and compare with the existing partial classification.
We conjecture that there are exactly $15$ five-dimensional reflexive polytopes that have $h^{1,1} = 1$.
We also present an example of a targeted search where conditions are placed on the Euler number of the CY manifold.
In Section~\ref{sec:conclusion}, we finish with a discussion and prospectus.
Our code, along with a database of five-dimensional reflexive polytopes we have generated is available on \href{https://github.com/elliheyes/Polytope-Generation}{\textsf{GitHub}}~\cite{githubcGA,Reflexive_Polytope_Genetic}.

\section{Background}\label{sec:background}
In this section we briefly review the necessary background, both on the mathematics of lattice polytopes and on genetic algorithms, while leaving some of the technical details to the appendices.

\subsection{Reflexive polytopes}\label{sec:refpol}
Due to theorems of Batyrev and Borisov~\cite{batyrev1993dual,batyrev1994calabiyau} reflexive polytopes provide an efficient way to construct Calabi--Yau (CY) manifolds.
(See Appendix~\ref{sec:appa} for a short review.)
This close connection to CY manifolds is the principal reason why physicists are interested in reflexive polytopes, and it prompted Kreuzer and Skarke to perform a \textit{tour de force} computer classification~\cite{Kreuzer_1997,kreuzer1998classification,kreuzer2000complete} which produced the largest available databases of smooth, compact CY manifolds in complex dimensions two and three.

Let us briefly review some of the properties of lattice polytopes relevant to our work.
An $n$-dimensional \emph{lattice polytope} $\Delta$ is the convex hull in $\mathbb{R}^n$ of a finite number of lattice points $x_1,\ldots ,x_m\in\mathbb{Z}^n\subset\mathbb{R}^n$.
These points can be conveniently combined into an $n \times m$ matrix $\cX=(x_1,\ldots ,x_{m})$ whose columns are the generators.
The vertices $v_1,\ldots ,v_{N_\text{v}}$ of $\Delta$ are a subset of the  lattice points $x_{i}$, so that $N_\text{v}=N_\text{v}(\Delta)\leq m$.
The vertices can also be combined into an $n \times N_{v}$ \emph{vertex matrix} $V=(v_1,\ldots ,v_{N_v})$.
Let $H=\{x\in\mathbb{R}^n\,|\, u\cdot x=d\}$ be a hyperplane where $u\in\mathbb{Z}^n$ is a primitive lattice point and $d\in \mathbb{R}$.
Such a hyperplane is called \emph{valid} if the polytope $\Delta$ is contained in the associated negative half-space, that is, if $u\cdot x\leq d$ for all $x\in\Delta$.
A \emph{face} of $\Delta$ is the intersection of $\Delta$ with a valid hyperplane, and a \emph{facet} is a face of dimension $n-1$.
We denote the set of all facets by $F(\Delta)$ and for a facet $\varphi\in F(\Delta)$ with equation $u\cdot x=d$ (where $u$ is a primitive lattice point) the number $d=d_\Delta(\varphi)$ is called the \emph{lattice distance} of $\varphi$ from the origin.
It is also useful to introduce the notation $N_\text{p}(\Delta)$ for the number of lattice points in $\Delta$.

As explained in Appendix~\ref{sec:appa}, a lattice polytope  
is said to have the \emph{IP property} if the origin is its only interior lattice point.
Furthermore, $\Delta$ is called \emph{reflexive} if it has the IP property and if its dual polytope $\Delta^*$ is also a lattice polytope and has the IP property.
Equivalently, $\Delta$ is reflexive if and only if it has the IP property and if all its facets have lattice distance one, that is, if $d_\Delta(\varphi)=1$ for all $\varphi\in F(\Delta)$.
It is the latter characterization of reflexivity which we will use later in our definition of the fitness function.

We consider two polytopes $\Delta$ and $\tilde{\Delta}$ with the same number of vertices, $N_\text{v}(\Delta)=N_\text{v}(\tilde{\Delta})=N_\text{v}$, as equivalent if their vertices are related by a common integer linear transformation combined with a permutation.
In other words, $\Delta$ and $\tilde{\Delta}$ are equivalent if there exist an $N_\text{v}\times N_\text{v}$ permutation matrix $P$ and a $G\in\text{GL}(n,\mathbb{Z})$ such that their vertex matrices $V$ and $\tilde{V}$ are related by
\begin{equation}\label{GVP}
     \tilde{V}=G\,VP\; .
\end{equation}
The most efficient way to eliminate the redundancy due to this identification is to define a normal form for the vertex matrix, thereby selecting precisely one representative per equivalence class.
The definition of this normal form and an algorithm for its computation is reviewed in Appendix~\ref{sec:NF}.
It is known that the number of reflexive polytopes, after modding out the identification~\eqref{GVP}, is finite in any given dimension $n$~\cite{lagarias_ziegler_1991}.
The connection between reflexive polytopes and CYs is further discussed in Appendix~\ref{sec:appa}.

\subsection{Genetic algorithms}
\label{sec:GA}
Genetic algorithms (GAs) are optimization algorithms which mimic the process of natural selection~\cite{darwin}.
They were first put forward in the 1950s~\cite{turing,barricelli} and were later formalized by Holland~\cite{Holland1975}.
Some more recent reviews are~\cite{David1989,Holland1992,Reeves2002,Charbonneau:2002,haupt,Michalewicz2004}.

GAs operate on a certain state space called the environment, which is frequently (and indeed in our applications) taken to be the set $\mathbb{F}^{n_\text{bits}}_2$ which consists of all bit lists with length $n_\text{bits}$. The elements of this set are often referred to as genotypes. 
In our case the environment consists of lattice polytopes $\Delta$ in $n$ dimensions which are generated as the convex hull of $m$ vectors $x_{a}\in\mathbb{Z}^{n}$, where $a=1,...,m$. These vectors are arranged into an $n\times m$ matrix $\cX=(x_1,\ldots ,x_m)$.
A detailed explanation of the environment and how the matrices $\cX$ are converted into a bitlists is given in Appendix~\ref{sec:env}. 

Further, we have two given functions, a fitness function $f:\mathbb{F}^{n_\text{bits}}_2\rightarrow \mathbb{R}$ whose value the algorithm is attempting to optimize and a probability distribution $p_\text{in}:\mathbb{F}^{n_\text{bits}}_2\rightarrow [0,1]$, which is used to select the initial population.
Given the bijection between the spaces of genotypes and phenotypes these functions can be defined on either space and we opt for the latter.
For the sampling probability $p_\text{in}$ it is usually sufficient to use a flat distribution, that is, every state in the state space has the same probability.\footnote{For large ranges of the matrix entries it can be advantageous to choose a non-flat $p_\text{in}$ which favors the selection of $x_a^i$ with smaller $|x_a^i|$.}
In our case the basic fitness function is defined as
\begin{equation}\label{ff}
    f(\Delta)=w_1 \left(\text{IP}(\Delta)-1\right)-\frac{w_2}{|F(\Delta)|}\sum_{\varphi\in F(\Delta)} |d_\Delta(\varphi)-1|\; ,
\end{equation}
where $\text{IP}(\Delta)$ equals $1$ if $\Delta$ has the IP property and is $0$ otherwise.
The numbers $w_1,w_2\in\mathbb{R}^{>0}$ are weights which are typically chosen as $w_1=w_2=1$.
Note that $f(\Delta)\leq 0$ always and $f(\Delta)=0$ if and only if $\Delta$ is reflexive.
Accordingly, we set $f_\text{term}=0$ so that the terminal states correspond to reflexive polytopes.

For some of our applications we are interested in a more targeted search for reflexive polytopes with certain additional properties.
For example, we might be interested in reflexive polytopes $\Delta$ whose number $N_\text{p}(\Delta)$ of lattice points equals a certain target $N_{\text{p},0}$.
Another interesting subclass of reflexive polytopes are those whose number of vertices $N_\text{v}(\Delta)$ matches a target $N_{\text{v},0}$.\footnote{Note that the number of vertices can be smaller than the number of generators $m$ that we start with, since generators can arise with multiplicity greater than one or can be contained in the interior of faces of $\Delta$.}
To facilitate such targeted searches, we can modify the fitness function~\eqref{ff} to
\begin{equation}
     \tilde{f}(\Delta)=f(\Delta)-w_3\,|N_\text{p}(\Delta)-N_{\text{p},0}|-w_4\, |N_\text{v}(\Delta)-N_{\text{v},0}|\; ,
     \label{eq:modified}
\end{equation}
where $w_3,w_4\in\mathbb{R}^{\geq 0}$ are two further weights which can be used to switch the additional requirements on and off.
If $w_3,w_4>0$, then $\tilde{f}(\Delta)=0$ and, hence, $\Delta$ is terminal, if and only if $\Delta$ is reflexive and has the target numbers $N_{\text{p},0}$ and $N_{\text{v},0}$ of lattice points and vertices.

The first step of a GA evolution is to select an initial population $P_0$ which contains a certain number, $n_\text{pop}$, of bit strings, each of length $n_\text{bits}$, by sampling the set $\mathbb{F}^{n_\text{bits}}_2$ with probability $p_\text{in}$.
The genetic evolution then consists of a sequence 
\begin{equation}
    P_0\rightarrow P_1\rightarrow \cdots\cdots\rightarrow P_{n_\text{gen}-1}\rightarrow P_{n_\text{gen}}
\end{equation}
of further $n_\text{gen}$ populations, each with the same size $n_\text{pop}$.
The basic evolutionary process, $P_k\rightarrow P_{k+1}$, to obtain population $k+1$ from population $k$ is carried out in three steps, namely, (i) selection, (ii) cross-over, and (iii) mutation.
We describe these three steps in turn.
\begin{itemize}
    \item [\textbf{(i)}] \textbf{Selection:}
    A probability distribution $p_k:P_k\rightarrow [0,1]$, based on the fitness function, is computed for the $k^\text{th}$ population.
    There are several ways to do this but the method we will employ here is the so-called roulette wheel selection where $p_k$ for an individual $s\in P_k$ is defined by
    \begin{equation}\label{pi}
         p_k(s)=\frac{1}{n_\text{pop}}\frac{(\alpha-1)\left(f(s)-\bar{f}\right)+f_\text{max}-\bar{f}}{f_\text{max}-\bar{f}}\; ,
    \end{equation}
    where $\bar{f}$ and $f_\text{max}$ are the average and maximal fitness values on $P_k$, respectively.
    The parameter $\alpha$, typically chosen in the range $\alpha\in[2,5]$, indicates by which factor the fittest individual in the population is more likely to be selected than the average one.
    Based on this probability $p_k$, $n_\text{pop}/2$ pairs are selected from the population $P_k$.
    \item [\textbf{(ii)}] \textbf{Cross-over:}
    For each pair selected in step (i), a random location $l\in\{1,\ldots ,n_\text{bits}\}$ along the bit string is chosen and the tails of the two strings are swapped.
    Carrying this out for all pairs leads to $n_\text{pop}$ new bit strings, which form the precursor, $\tilde{P}_{k+1}$, of the new population.
    \item [\textbf{(iii)}] \textbf{Mutation:}
    In the final step, a certain fraction, $r_\text{mut}$, of bits in the population $\tilde{P}_{k+1}$ from step (ii) is flipped and this produces the next generation $P_{k+1}$.
\end{itemize}
A common addition to the above algorithm (which we employ in our applications) is \emph{elitism} which means that the fittest individual from population $P_k$ is copied to the population $P_{k+1}$ unchanged.
In summary, a GA evolution is subject to the following hyper-parameter choices: the population size $n_\text{pop}$, the number of generations $n_\text{gen}$, the parameter $\alpha$ in~\eqref{pi} and the mutation rate $r_\text{mut}$.
For a systematic search, typically many GA evolutions, each with a new randomly sampled initial population $P_0$, are carried out.
Then, the desirable (or ``terminal") states $s$, defined as states with $f(s)\geq f_\text{term}$ for a certain critical value $f_\text{term}$, are extracted from all populations which arise in this way.

Once the GA has found a list of reflexive polytopes, possibly with additional properties, we are not yet finished, since we have to eliminate the redundancies which arise from the identification~\eqref{GVP}.
This is done by computing the normal form of the vertex matrix, using the algorithm described in Appendix~\ref{sec:NF}.

For our applications, we use \textsf{PALP}~\cite{Kreuzer_2004} tools for polytope computation and additional \textsf{c} code by the authors~\cite{Reflexive_Polytope_Genetic} which realizes the phenotype-genotype conversion. This is combined into a lightweight and fast \textsf{c} code~\cite{githubcGA} which realizes the GA.

\section{Low dimensional results}\label{sec:lower_dim_results}
To showcase the capability of GAs for searching reflexive polytopes, we start with $n=2,3,4$ dimensions where complete classification already exists.\footnote{
The one-dimensional classification consists of a single reflexive polytope formed from two integer points $\pm 1$ adjacent to the origin.
Since there is only one polytope of this type, we ignore this dimension.} As we will see, these results provide useful guidance for the search in $n=5$ dimensions where a complete classification is lacking.

There are some common hyperparameter choices which we use for all following runs (including those in Section \ref{sec:5d_results}).
In each case, we evolve populations for $n_\text{gen}=500$ generations, we use a mutation rate of $r_\text{mut}=0.005$, and the parameter $\alpha$ in~\eqref{pi} is set to $\alpha=3$.
Other hyperparameters, such as the population size $n_\text{pop}$, and environmental variables will be chosen to optimize results and their values for each case will be stated below.

\subsection{Two and three dimensions}\label{sec:2d+3d}
In two dimensions, where there are 16 unique reflexive polytopes, we use an integer range $x_a^i\in [-3,4]$, so that each integer is encoded by $\nu=3$ bits, and $m=6$ generators.
Hence, each matrix $\cX$ is represented by $n_\text{bits}=36$ bits and the  environment consists of $10^{36}\simeq 10^{11}$ states.

The total number of reflexive polytope states in the environment will of course be greater than 16 because equivalence relations exist by the identification~\eqref{GVP}. We'd like to estimate the number of such states in the environment, the details of how we compute this estimate are given in Appendix \ref{sec:GLnZ}. In short, we estimate the number $N$ of $\text{GL}(2,\mathbb{Z})$ matrices which transform the unique reflexive polytopes into equivalent reflexive polytopes that exist within the environment, i.e. have vertex coordinates in the range $[-3,4]$. The estimate on the number of reflexive polytopes in the environment is then given as (\# unique reflexive polytopes)$\times N$. We estimate $N$ to be $107$, and therefore we estimate that there are $16\times107\sim10^{3}$ number of reflexive polytopes in the environment.

Using a population size of $n_\text{pop}=200$, the genetic algorithm finds all $16$ reflexive polytopes after only one evolution, taking only a few seconds on a single CPU.
Assuming that the GA never visits the same state twice, the total number of states visited would be $\text{\# evolutions}\times n_{\text{gen}}\times n_{\text{pop}} = 1\times500\times200 \simeq 10^{5}$ which is only a fraction of $\sim 10^{-6}$ of the total environmental states. In reality, the GA is likely to visit states more than once so this is in fact an upper bound and the true fraction of states visited will be smaller. 

In three dimensions, we use the coordinate range $x_a^i\in [-7,8]$, with $\nu=4$ bits per integer and $m=14$ generators.
This means each matrix $\cX$ is described by $n_\text{bits}=168$ bits and the total environment size is $2^{168}\simeq 10^{51}$.
With population size set to $n_\text{pop}=450$, the genetic algorithm finds all $4319$ reflexive polytopes after $117251$ evolutions. The upper bound of states visited by the GA is $117251 \times 500 \times 450 \simeq 10^{11}$, which is a very small fraction of $\sim 10^{-40}$ of the environmental states. For comparison, we estimate the total number of reflexive polytopes in the environment to be $\sim 10^{9}$ ($\sim 10^{-42}$ of the total environment). Considering the small fraction of reflexive polytopes in the environment this it is a remarkable achievement that the GA managed to find all the unique ones.  

\subsection{Four dimensions}\label{sec:4d}
In four dimensions, the total number  of reflexive polytopes is large ($473,800,776$) and, for our purpose of obtaining baseline performance results to inform the five-dimensional search later on, it is not necessary to generate the complete set.
Instead, we focus our attention on finding those polytopes $\Delta$ with the lowest number of vertices and points.
In $n=4$ dimensions the minimum number of vertices is $5$ and therefore the minimum number of points (assuming reflexive polytopes for which the origin is the single interior point) is $6$.
To facilitate such a search we use the modified fitness function~\eqref{eq:modified} with certain targets $N_{\text{v},0}$ or $N_{\text{p},0}$ for the number of vertices or points.

\begin{figure}[h]
    \centering
    \includegraphics[width=0.6\textwidth]{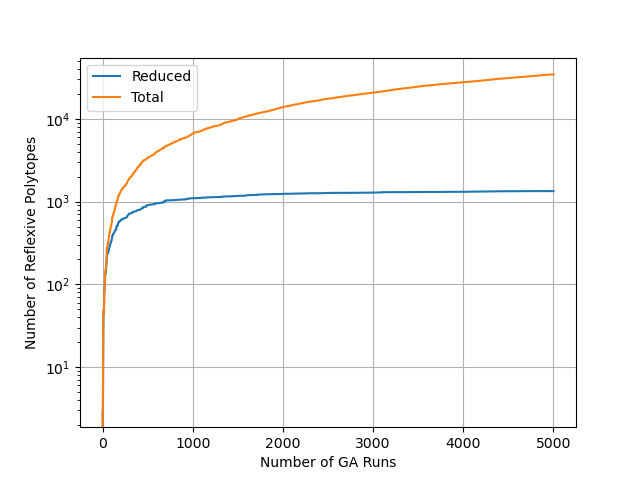}
    \caption{\sf Log plot of total number of generated four-dimensional reflexive polytopes with five vertices against number of genetic algorithm evolutions.
    The total before and after removing redundancy are shown in orange and blue, respectively.}
    \label{fig:4d_5v}
\end{figure}

We first perform a search for reflexive polytopes with the lowest number, $N_{\text{v},0}=5$, of vertices (and an arbitrary number of points).
This means we set $w_3>0$ and $w_4=0$ in the modified fitness function~\eqref{eq:modified}.
The integer range is taken to be $x_a^i\in [-15,16]$, that is, we use $\nu=5$ bits per integer and $m=5$ generators.
This means the matrices $\cX$ are described by bit strings of length $n_\text{bits}=100$ and the environment contains $2^{100}\simeq 10^{30}$ states.
Furthermore, we estimate the number of reflexive polytopes in this environment to be $\sim 10^{15}$, which is still a very small fraction $\sim 10^{-15}$ of the total environment.
The population size is taken to be $n_\text{pop}=200$.
With these settings we performed multiple GA runs and the number of reflexive polytopes as a number of generations obtained is shown in Figure~\ref{fig:4d_5v}.

Evidently, the number of reflexive polytopes, after removing the redundancy due to~\eqref{GVP}, saturates quickly and to a value of $1555$, just six reflexive polytopes short of the total of $1561$.
The missing six polytopes all have Euler number $\chi=0$ and large vertex coefficients in their normal form.
Applying a few million $\text{GL}(4,\mathbb{Z})$ transformations, we are unable to find a single equivalent vertex matrix for these six cases which falls into our integer range.
This suggests these cases can only be found by enlarging the integer range $x_a^i\in [-15,16]$.
We will refrain from doing this as the present run has already found $99.6\%$ of all reflexive polytopes and provides strong evidence that, given appropriate hyperparameter and environmental choices, GAs can find virtually complete sets of reflexive polytopes.

Next, we are searching for polytopes with a given number, $N_{\text{p},0}$ of lattice points, so we set $w_3=0$ and $w_4>0$ in the fitness function~\eqref{eq:modified}.
Specifically, we will be focusing on the cases $N_{\text{p},0}\in\{6,7,8,9,10\}$.
The vertex coefficients of such polytopes with a relatively small number of points are likely to be small.
Therefore, we reduce the integer range to $x_a^i\in [-3,4]$ and describe every integer by $\nu=3$ bits.
This leads to a reduction of the environment size and a significant improvement in the algorithm's performance, compared to the previous case. If a polytope has $N_{p,0}$ points, the maximum number of vertices is $N_\text{v}=N_{\text{p},0}-1$, where all points are vertices except the origin. Therefore, in searching for polytopes with $N_{\text{p},0}$ points we set the number of generators to $m=N_{\text{p},0}-1$. 
The results are summarised in Table~\ref{tab:4d_Np}.
\begin{table}[h!]
\centering
\addtolength{\leftskip}{-1.75cm}
\addtolength{\rightskip}{-1.75cm}
 \begin{tabular}{|| c c c c c c c ||} 
 \hline
 \# points & \# states & \# un. refl.~poly. & \# refl.~poly. & $n_{\text{pop}}$ & \# GA runs & \% states visited\\ [0.5ex] 
 \hline\hline
 $6$ & $\sim10^{19}$ & $3$ & $\sim10^{10}$ & $400$ & $5$ & $\sim10^{-13}$ \\ 
 $7$ & $\sim10^{22}$ & $25$ & $\sim10^{11}$ & $300$ & $30$ & $\sim10^{-16}$ \\
 $8$ & $\sim10^{26}$ & $168$ & $\sim10^{12}$ & $400$ & $60$ & $\sim10^{-19}$ \\
 $9$ & $\sim10^{29}$ & $892$ & $\sim10^{12}$ & $300$ & $9378$ & $\sim10^{-20}$ \\
 $10$ & $\sim10^{33}$ & $3838$ & $\sim10^{13}$ & $350$ & $9593$ & $\sim10^{-24}$ \\ [1ex] 
 \hline
 \end{tabular}
 \caption{\sf Results for four-dimensional reflexive polytopes with the a small number of lattice points, as in the first column. The size of the environment used for the GA is provided in the second column, the total number of unique reflexive polytopes for the given number of points (as taken from the Kreuzer--Skarke list) is given in the third column, the fourth column gives an estimate on the total number of reflexive polytopes in the environment, in the fifth column is the population size, the sixth column gives the number of GA runs required to find all unique reflexive polytopes and the last column provides an upper bound on the fraction of states visiting during all GA runs.}
\label{tab:4d_Np}
\end{table}
It is remarkable that all states are found in all cases after a sufficient number of GA runs.

\section{Five-dimensional results}\label{sec:5d_results}
In the previous section, we have seen that GAs can generate complete or near-complete lists of reflexive polytopes in two, three, and four dimensions. This is a valuable proof of principle which demonstrates that GAs can successfully identify reflexive polytopes. However, the results are of limited practical use, given the complete classifications in those dimensions. We now  turn to reflexive polytopes in five dimensions, the lowest-dimensional case for which a complete classification is not available.
The total number of (inequivalent) reflexive polytopes in dimensions $n=1,2,3,4$ is given by $1$, $16$, $4319$, $473,800,776$, respectively. This sequence suggests the number of reflexive polytopes in five dimensions is extremely large and producing a complete catalog is intractable.\footnote{Extrapolating this trend gives an estimate of $1.15 \times 10^{18}$ five-dimensional reflexive polytopes~\cite{Sch_ller_2019}.}
The partial list of Sch\"oller and Skarke of $185,269,499,015$ weight systems that give rise to (not necessarily inequivalent) maximal five-dimensional reflexive polytopes~\cite{Sch_ller_2019} is the state of the art.
The GA, however, is not biased towards generating maximal polytopes, and can be used to search for polytopes with other properties. In fact, it is likely that the vertices of the largest polytopes are far from the origin, and the GA would struggle to find such cases. For this reasons, we focus on generating small polytopes with a small number of points and vertices. We note that the ``small'' polytopes we generate are not necessarily dual to Skarke and Scholler’s ``maximal'' polytopes. The maximal polytopes in Skarke and Scholler’s list are those that contain every reflexive polytope as a subpolytope and so the dual of these will indeed have a small number of points. However, some of our small polytopes which have a small number of points might have a dual that is large (meaning it has large number of points) but it is still a subpolytope of an even larger maximal polytope, in which case the dual would not be in Skarke and Scholler’s database.

We can compute the normal forms of the polytopes found by the GA with those of the polytopes in the existing list, to confirm that we have indeed found new five-dimensional reflexive polytopes.

\subsection[$N_\text{v}=6$]{$\bm{N_\text{v}=6}$}\label{sec:5d_Nv}
\begin{figure}[h]
    \centering
    \includegraphics[width=0.6\textwidth]{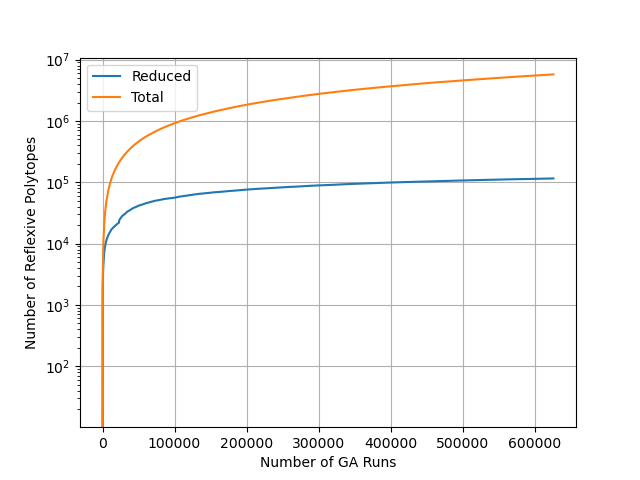}
    \caption{\sf Total number of generated five-dimensional reflexive polytopes with six vertices against number of genetic algorithm evolutions.}
    \label{fig:5d_6v}
\end{figure}

In analogy with the four-dimensional case, we start by looking at polytopes with the smallest number of vertices, that is, $N_\text{v}=6$.
We use the integer range $\{-15,\ldots, 16\}$ so that every integer is represented by $\nu=5$ bits and the total length of the genotype is $n_\text{bits}=150$. Hence, the size of the environment is $2^{150} \simeq 10^{46}$. We have performed $626318$ GA runs, taking over two months to complete on a single CPU.
We used $n_\text{gen}=500$ generations and a population size $n_\text{pop}=500$, meaning during the search a fraction of at most $10^{-35}$ of the environment has been visited. 
The number of reflexive polytopes against the number of runs found in this way is shown in Figure~\ref{fig:5d_6v}. After removing redundancies, we end up with a total of $115567$ five-dimensional reflexive polytopes with six vertices. From Figure~\ref{fig:5d_6v} we see that the the reduced number of reflexive polytopes is still increasing and therefore running the GA for longer we should find more six-vertex polytopes. However, due to limited resources and so we do not pursue this. 

\subsection[$N_\text{p}=7,8,9,10,11$]{$\bm{N_\text{p}=7,8,9,10,11}$}\label{sec:5d_Np}
We have also searched for those five-dimensional reflexive polytopes with the lowest number of points, that is, $N_{{\rm p},0}\in\{7,8,9,10,11\}$. This has been done for the range integer range $\{-3,\ldots, 4\}$, so that every integer is represented by $\nu=3$ bits, leading the bit strings of length $n_\text{bits}=15 (N_{{\rm p},0}-1)$. In each case, we perform as many GA runs as necessary until no new polytopes are found for $1000$ evolutions. The results are presented in Table~\ref{tab:5d_Np}.

\begin{table}[h!]
\centering
 \begin{tabular}{|| c c c c c||} 
 \hline
 \# points & \# states & $n_{\text{pop}}$ & \# un. refl.~poly.& \# GA runs\\ [0.5ex] 
 \hline\hline
 $7$ & $\sim10^{28}$ & $350$ & $9$ & $36$ \\ 
 $8$ & $\sim10^{32}$ & $350$ & $115$ & $1278$ \\
 $9$ & $\sim10^{37}$ & $450$ & $1385$ & $7520$ \\
 $10$ & $\sim10^{41}$ & $750$ & $12661$ & $31857$ \\
 $11$ & $\sim10^{46}$ & $650$ & $94556$ & $376757$ \\ [1ex] 
 \hline
 \end{tabular}
 \caption{\sf Results for five-dimensional reflexive polytopes with the a small number of lattice points, as in the first column. The size of the environment used for the GA is provided in the second column, the population size in the third column, the fourth column lists the total number of unique reflexive polytopes found by the GA for the given number of points and the last column gives the number of GA runs at which the total list of reflexive polytopes saturates and after which no new reflexive polytopes are found for 1000 runs.}
\label{tab:5d_Np}
\end{table}

Of course we do not know with certainty which fraction of low-point polytopes we have found in this way. It is possible that some five-dimensional reflexive polytopes with such numbers of points still exist. On the other hand, in view of the highly successful low-point searches in four dimensions, it seems likely we have found a large fraction of those polytopes. 

\subsection[$h^{1,1}=1,2$]{$\bm{h^{1,1}=1,2}$}\label{sec:5d_h11}
It is interesting to ask about CY manifolds with small Hodge numbers and here we focus on cases with small $h^{1,1}$. The Sch\"oller--Skarke list contains $8$ weight systems corresponding to CY hypersurfaces with $h^{1,1}=1$ and $33$ weight systems corresponding to CY hypersurfaces with $h^{1,1}=2$. In their list there are also $8,409,140$ and  $186,659,154$ weight systems with $h^{1,3}=1,2$ respectively, whose duals will have $h^{1,1}=1,2$. Taking a sample of $100,000$ weight systems in each case and computing the corresponding reflexive polytopes in their normal forms we find that there are only $7$ and $47$ unique polytopes. This is a huge reduction and highlights the large amount of redundancy in the Sch\"oller--Skarke list of weight systems. 
We have scanned the lists of five-dimensional reflexive polytopes obtained from the GA runs described above and have found $15$ polytopes with $h^{1,3}=1$ whose dual polytopes have $h^{1,1}=1$. Similarly, we have found $195$ polytopes with $h^{1,3}=2$ whose dual polytopes have $h^{1,1}=2$. By comparing normal forms, we have verified that these dual polytopes contain all $h^{1,1}=1,2$ polytopes from the Sch\"oller--Skarke list. In addition, there are many new examples, even some with Hodge numbers which are not contained in the Sch\"oller--Skarke list or in the list of four-dimensional CY manifolds realized as complete intersections in products of projective space (CICYs)~\cite{Gray:2013mja, cicy4}. Two such examples with a new set of Hodge numbers are given in Appendix~\ref{sec:examples}.

All $15$ polytopes with $h^{1,3}=1$ arise from the datasets with $N_{\text{p}}=7,8$ and no such polytopes are found for $8<N_\text{p}\leq 11$. We take this as evidence that the dataset is complete.

\begin{conjecture}
    There are precisely $15$ inequivalent (non-isomorphic) five-dimensional reflexive polytopes that give rise to four complex dimensional Calabi--Yau hypersurfaces with Hodge number $h^{1,1}=1$.
\end{conjecture}

\subsection{Targeted searches}\label{sec:targeted_search}
To showcase the capability of our GA at generating CY four-folds with specific criteria, we present an example of a targeted search inspired by~\cite{Berg_2003}.
In that paper, the authors consider eleven-dimensional supergravity compactified on CY four-folds with $4$-form flux and provide the conditions necessary to break supersymmetry from $\mathcal{N}=2$ to $\mathcal{N}=1$.
In Appendix~A they search for CY four-folds with $\chi$ divisible by $\delta\in\{24,224,504\}$ which satisfy the $\mathcal{N}=1$ condition.
By searching the Sch\"oller--Skarke list they find eight examples which they present in their Table~1. To facilitate a GA search for such cases, we modify our fitness function to
\begin{equation}
     \tilde{f}(\Delta)=f(\Delta)- w_5 \sum_{\delta} \chi(\Delta)\mbox{ mod }\delta \; ,
\end{equation}
where $w_5$ is a weight and $\chi(\Delta)$ is the Euler number of $\Delta$. 
In our search for such polytopes we set the number of generators to be $m=10$ and use the integer coordinate range $[-3,4]$, where each integer is represented by a $\nu=3$ bits. 
With population size $n_\text{pop}=550$ and  after $10000$ evolutions the GA finds $1871$ polytopes that satisfy the index condition and, comparing the Hodge numbers with those in Sch\"oller and Skarke's list, we find that two of these are new. One such example is given in Appendix~\ref{sec:examples}.

This example illustrates the possibilities of a targeted GA search. By a suitable modification of the fitness function one can design a dedicated search for CY manifolds with prescribed properties, for example with certain values of the Euler number as above, but also with a given pattern of Hodge numbers, with Chern classes and intersection form satisfying certain constraints or combinations of all of these. This points to a different approach for dealing with large classes of geometries in string theory. Rather than producing complete lists of such geometries (which is not even feasible for the case at hand, that is, five-dimensional reflexive polytopes) the GA can be used to search for geometries with prescribed properties, as required for the intended string compactification.

\section{Discussion}\label{sec:conclusion}
In this paper we have shown that genetic algorithms (GAs) can be efficiently used to generate reflexive polytopes in two, three, four, and five dimensions.
In two dimensions we have generated the complete set of $16$ reflexive polytopes in just one GA evolution. 
We have also generated the complete set of reflexive polytopes in three dimensions in $\sim 100000$ evolutions.
Due to the large number of reflexive polytopes in four dimensions, we have refrained from generating the complete set in this case. Instead, we have focused on the polytopes with the smallest number $N_{{\rm p},0}$, of lattice points, that is, $6\leq N_{{\rm p},0}\leq 10$. By comparing with the Kreuzer--Skarke classification, we have shown that the GA can find all such polytopes. These results indicate that complete or near-complete classifications of reflexive polytopes can be accomplished with GAs, at least for cases with a small number of lattice points.

This observation is important for the five-dimensional case, where only a partial classification of reflexive polytopes exists. Performing a GA search for $7\leq N_{{\rm p},0}\leq 11$ in five dimensions produces all reflexive polytopes from the partial classification and indeed many more, previously unknown cases. This includes cases which lead to CY four-folds with new sets of Hodge numbers. While the numbers of reflexive polytopes obtained in this way (see Table~\ref{tab:5d_Np}) is unlikely to be the true total there are good indications that they provide strong lower bounds. From these lists, we have also extracted all polytopes with $h^{1,1}=1$. We conjecture that the $15$ cases found constitute the complete list of reflexive polytopes which give rise to CY four-folds with $h^{1,1}=1$.

It is perhaps not desirable, or even feasible, to generate the complete list of reflexive polytopes beyond four dimensions. Instead, we propose an alternative approach, well-suited to the needs of string compactifications, of targeted searches for reflexive polytopes (and their associated CY manifolds) with certain prescribed properties. We have demonstrated that GAs can be used for such targeted searches, by looking for cases with certain prescribed values of the Euler number. This has led to new reflexive polytopes that satisfy the condition for M-theory compactifications on CY four-folds, following~\cite{Berg_2003}. We expect the same approach will work for other targets, such as a certain desirable pattern of Hodge numbers. 

The \textsf{c} code underlying the above results and all data sets are available on \href{https://github.com/elliheyes/Polytope-Generation}{\textsf{GitHub}}~\cite{Reflexive_Polytope_Genetic,githubcGA}.

There are many possible directions for future research.
In particular, by fine, star, regular triangulation of a (dual) reflexive polytope into simplices, we can construct the CY hypersurface explicitly.
This process is also amenable to attack with GAs.
Targeted GA searches are another promising avenue.
For example, it might be possible to design a targeted search for elliptically or K3 fibered CY four-folds. More ambitiously, one can aim for searches which produce F-theory compactifications with certain desirable properties. It might also be interesting to apply reinforcement learning to the problem of searching for reflexive polytopes and compare its performance to that of GAs. We leave this to future work.

\section*{Acknowledgements}
We thank the organizers of the ``Deep Learning Era of Particle Theory'' 2022 conference at the Mainz Institute for Theoretical Physics, where this collaboration was initiated. PB, E.~Heyes, and E.~Hirst would also like to thank Pollica Physics Center, Italy, and the participants at the workshop ``At the Interface of Physics, Mathematics, and Artificial Intelligence'' for a very stimulating environment at the completion of this work.
PB is supported in part by the Department of Energy grant DE-SC0020220.
YHH is supported by STFC grant ST/J00037X/2.
E.~Heyes is supported by City, University of London and the States of Jersey.
E.~Hirst is supported by Pierre Andurand.
VJ is supported by the South African Research Chairs Initiative of the Department of Science and Innovation and the National Research Foundation.

\appendix

\section{Calabi--Yau manifolds from reflexive polytopes}\label{sec:appa}
In this section, we briefly review the necessary elements of toric geometry, with the goal of introducing the construction of mirror pairs of Calabi--Yau (CY) manifolds from reflexive polytopes. 

\begin{definition} 
    Let $M \cong \mathbb{Z}^{n}$ and $N = \text{Hom}(M,\mathbb{Z})$ be a dual pair of lattices with the pairing $\langle \cdot,\cdot \rangle: N \times M \rightarrow \mathbb{Z}$, and let $M_{\mathbb{Q}}$, $N_{\mathbb{Q}}$ be their rational extensions.
    \begin{itemize}
        \item A polytope $\Delta$ in $M_{\mathbb{Q}}$ is the convex hull of finite number of points in $M_{\mathbb{Q}}$.
        \item $\Delta$ is called a lattice polytope if all its vertices lie in $M$.
        \item The dual or polar polytope of $\Delta$ is defined as
        \begin{equation}
            \Delta^{*} = \{ n \in N_{\mathbb{Q}} | \langle n,m \rangle \geq -1\ \forall\, m \in \Delta \} ~.
        \end{equation}
        \item A face $\theta$ of $\Delta$ is defined as
        \begin{equation}
            \theta = \{ m \in \Delta | \langle n,m \rangle = r \} ~,
        \end{equation}
        for some $n \in N_{\mathbb{Q}}$ and $r \in \mathbb{R}$.
    \end{itemize}
\end{definition}

Given an $n$-dimensional lattice polytope $\Delta$, one can construct a compact toric variety $X_{\Delta}$ of complex dimension $n$.
In short, one constructs the normal fan $\Sigma_{\Delta}$ as follows: for a face $\theta$ of $\Delta$, let $\sigma_{\theta} \subset N_{\mathbb{R}}$ be the dual of the cone:
\begin{equation}
    \sigma_{\theta}^{\vee} := \{ \lambda(u-u') | u \in \Delta, u' \in \theta, \lambda \geq 0 \} \subset M_{\mathbb{R}} ~.
\end{equation}
Then the normal fan is given as $\Sigma_{\Delta}:=\{\sigma_{\theta}\}$ for all faces $\theta$ of $\Delta$.
From the normal fan, the construction of the compact toric variety $X_{\Delta}$ follows the usual procedure~\cite{fulton}, where each cone gives rise to an affine toric variety and one glues these patches together. 

\begin{definition}
    A polytope is said to satisfy the interior point (IP) property when it contains only one interior point taken to be the origin.
    Let $\Delta \subset M$ be a lattice polytope satisfying the IP property, then $\Delta$ is called reflexive if its dual $\Delta^{*} \subset N$ is also a lattice polytope satisfying the IP property.
\end{definition}

We recall that a CY $n$-fold $\mathcal{M}$ is an $n$ complex dimensional space that is a compact K\"ahler manifold and has a vanishing first real Chern class.
Calabi conjectured and Yau proved that such a geometry admits a unique Ricci-flat metric in each K\"ahler class.

The connection between CY manifolds and reflexive polytopes is the following.
Let $\Delta \subset M$ be an $n$-dimensional reflexive polytope and $X_{\Delta}$ the corresponding $n$ complex dimensional toric variety.
Then it follows that the zero locus of a generic section of the anticanonical bundle $-K_X$ is a CY variety $\mathcal{M}$ of dimension $n-1$ which can be resolved into a CY orbifold with at most terminal singularities.
The mirror CY $\mathcal{W}$ is similarly obtained from the polar dual.
(Because of the IP property, it turns out that $(\Delta^*)^* = \Delta$.)
See~\cite{Altman:2014bfa,Demirtas:2022hqf} for explicit constructions of CY manifolds from reflexive polytopes.

\section{Normal form}\label{sec:NF}
There are two sources of redundancy when defining a reflexive polytope $\Delta$ in $n$ dimensions by its $n \times N_\text{v}$ vertex matrix $V$, whose columns are the $N_\text{v}$ vertices.
First of all, one permute the vertices, leading to an $S_{N_\text{v}}$ symmetry which permutes the columns of $V$. Secondly, one can perform a coordinate transformation on the $n$-dimensional lattice by acting on $V$ with a ${\rm GL}(n,\mathbb{Z})$ matrix from the left. Altogether, this amounts to a transformation of the vertex matrix as in Eq.~\eqref{GVP}.

In order to remove the redundancy in the list of polytopes, we compare their normal forms.
This is the approach that was used by Kreuzer and Skarke in constructing the complete classification of three- and four-dimensional reflexive polytopes~\cite{kreuzer1998classification,kreuzer2000complete} and is included in the \textsf{PALP} software package~\cite{Kreuzer_2004}.
If two polytopes $\Delta_{1}$ and $\Delta_{2}$ have the same normal form, then they are equivalent, in the sense that they are isomorphic with respect to a lattice automorphism.
A detailed description of the how one computes the normal form is given in~\cite{grinis2013normal}.
We shall give a short description here. 

Let $M$ be a $n$-dimensional lattice and $\Delta \subset M_{\mathbb{Q}}$ a $n$-dimensional lattice polytope with $N_\text{v}$ vertices, $N_\text{f} = |F(\Delta)|$ facets and vertex matrix $V$.
We also define the supporting hyperplanes of $\Delta$, associated to the facets $\varphi_{i}\in F(\Delta)$, as the set of all vectors $v$ satisfying $\langle w_{i},v \rangle = -c_{i}$, where $(w_{i},c_{i}) \in M^{*} \times \mathbb{Z}$. The algorithm to compute the normal form is then as follows.

\begin{enumerate}
    \item Compute the $N_\text{f} \times N_\text{v}$ vertex-facet pairing matrix $PM$:
    \begin{equation}
        PM_{ij}:= \langle w_{i},v_{j} \rangle + c_{i} ~.
    \end{equation}
    \item Order the pairing matrix $PM$ lexicographically to get the maximal matrix $PM^{\text{max}}$.
    \item Further rearrange the columns of $PM^{\text{max}}$ to get $M$ by the following: 
    \begin{align*}
        & M \leftarrow PM^{\text{max}} \\
        & \textbf{for } i=1 \textbf{ to } N_\text{v} \textbf{ do} \\
        & \hspace{0.5cm} k \leftarrow i \\
        & \hspace{0.5cm} \textbf{for } j=i+1 \textbf{ to } N_\text{v} \textbf{ do} \\
        & \hspace{1cm} \textbf{if } c_{M}(j) < c_{M}(k) \vee (c_{M}(j)=c_{M}(k) \wedge s_{M}(j) < s_{M}(k)) \textbf{ then} \\
        &\hspace{1.5cm} k \leftarrow j \\
        & \hspace{1cm} \textbf{end if} \\
        & \hspace{0.5cm} \textbf{end for} \\
        & \hspace{0.5cm} M \leftarrow \text{SwapColumn}(M,i,k) \\
        & \textbf{end for}
    \end{align*}
    where $c_{M}(j):=\text{max}(M_{ij} |1 \leq i \leq N_\text{f})$ and $s_{M}(j):=\sum_{i=1}^{N_\text{f}}M_{ij}$, where $1\leq j\leq N_{\text{v}}$.
    \item Let $\sigma_{\text{max}}$ denote the associated element of $S_{N_\text{f}} \times S_{N_\text{v}}$ that transforms $PM$ into $M$. Order the columns of $V$ according to the restriction of  $\sigma_{\text{max}}$ to $S_{N_\text{v}}$ to get the maximal vertex matrix $V^{\text{max}}$. This removes the permutation degeneracy.
    \item Compute the row style Hermite normal form of $V^{\text{max}}$ to obtain the normal form $NF$. This step removes the $GL(n,\mathbb{Z})$ degeneracy.
\end{enumerate}

\begin{example}
    To illustrate the above algorithm, we present an example in three dimensions. Let $\Delta$ be a lattice polytope defined by the vertex matrix:
    \begin{equation}
        V = 
        \left(\begin{array}{rrrrrrrr}
            0 & -2 & -1 & -1 & 1 & -3 & 2 & -2 \\
            1 & -3 & -1 & 0 & 0 & -3 & 1 & -1 \\
            1 & -3 & -2 & -1 & 0 & -4 & 3 & -2
        \end{array}\right)\; .
    \end{equation}
    Computing the vertex-facet pairing matrix we get
    \begin{equation}
        PM = 
        \left(\begin{array}{rrrrrrrr}
            2 & 0 & 0 & 0 & 2 & 1 & 1 & 0 \\
            0 & 1 & 0 & 0 & 2 & 2 & 2 & 1 \\
            0 & 6 & 6 & 0 & 0 & 1 & 3 & 2 \\
            2 & 1 & 2 & 2 & 0 & 0 & 0 & 1 \\
            3 & 0 & 1 & 1 & 1 & 0 & 0 & 0 \\
            2 & 5 & 6 & 0 & 0 & 0 & 2 & 1 \\
            3 & 2 & 3 & 0 & 1 & 0 & 1 & 0 \\
            0 & 0 & 0 & 3 & 0 & 1 & 0 & 2 \\
        \end{array}\right)\; .
    \end{equation}
    Ordering $PM$ lexicographically we get the following maximal matrix:
    \begin{equation}
        PM^{\text{max}} = 
        \left(\begin{array}{rrrrrrrr}
            6 & 6 & 3 & 2 & 1 & 0 & 0 & 0 \\
            6 & 5 & 2 & 1 & 0 & 2 & 0 & 0 \\
            3 & 2 & 1 & 0 & 0 & 3 & 1 & 0 \\
            2 & 1 & 0 & 1 & 0 & 2 & 0 & 2 \\
            1 & 0 & 0 & 0 & 0 & 3 & 1 & 1 \\
            0 & 1 & 2 & 1 & 2 & 0 & 2 & 0 \\
            0 & 0 & 1 & 0 & 1 & 2 & 2 & 0 \\
            0 & 0 & 0 & 2 & 1 & 0 & 0 & 3 \\
        \end{array}\right)\; , 
    \end{equation}
    corresponding to the row and column permutations $(3,6,7,4,5,2,1,8)$ and $(3,2,7,8,6,1,5,4)$ respectively.
    Further ordering the columns by the procedure described in Step 3 above we get
    \begin{equation}
        M = 
        \left(\begin{array}{rrrrrrrr}
            0 & 1 & 0 & 2 & 0 & 6 & 6 & 3 \\
            2 & 0 & 0 & 1 & 0 & 6 & 5 & 2 \\
            3 & 0 & 0 & 0 & 1 & 3 & 2 & 1 \\
            2 & 0 & 2 & 1 & 0 & 2 & 1 & 0 \\
            3 & 0 & 1 & 0 & 1 & 1 & 0 & 0 \\
            0 & 2 & 0 & 1 & 2 & 0 & 1 & 2 \\
            2 & 1 & 0 & 0 & 2 & 0 & 0 & 1 \\
            0 & 1 & 3 & 2 & 0 & 0 & 0 & 0 \\
        \end{array}\right)\; ,
    \end{equation}
    corresponding to the column permutation $(6,5,8,4,7,1,2,3)$. Ordering the columns in $V$ correspondingly we get 
    \begin{equation}
        V^{\text{max}} = 
        \left(\begin{array}{rrrrrrrr}
            -3 & 1 & -2 & -1 & 2 & 0 & -2 & -1 \\
            -3 & 0 & -1 & 0 & 1 & 1 & -3 & -1 \\
            -4 & 0 & -2 & -1 & 3 & 1 & -3 & -2
        \end{array}\right)\; .
    \end{equation}
    Finally, computing the row style Hermite normal form of $V^{\text{max}}$ we arrive at the following normal form:
    \begin{equation}
        NF = 
        \left(\begin{array}{rrrrrrrr}
            1 & 0 & 1 & 0 & 1 & -2 & 1 & 0 \\
            0 & 1 & -1 & 0 & -1 & 1 & -3 & -3 \\
            0 & 0 & 0 & 1 & -1 & 0 & -2 & -2
        \end{array}\right)\; .    
    \end{equation}
\end{example}

\section{Environment}\label{sec:env}
In order to set up the environment, we consider lattice polytopes $\Delta$ in $n$ dimensions which are generated as the convex hull of $m$ vectors $x_a\in\mathbb{Z}^n$, where $a=1,\ldots ,m$.
These vectors are arranged into an $n\times m$ matrix $\cX=(x_1,\ldots ,x_m)$.

In practice, we must restrict the entries $x_a^i$ of the matrix $\cX$ to a finite range which we choose to be $x_a^i\in\{x_\text{min},x_\text{min}+1,\ldots, x_\text{min}+2^\nu-1\}$, for certain integers $x_\text{min}$ and $\nu$.
Our environment $E$ therefore consists of all $n\times m$ integer matrices $\cX$ with entries in this range.
The elements of an environment $E$ are also referred to as phenotypes.
The first step in applying GAs to such an environment is to define the phenotype-genotype map $E\rightarrow \mathbb{F}^{n_\text{bits}}_2$.
Given our choices this is quite straightforward.
Each integer $x_a^i$ is converted into a bit string of length $\nu$ and concatenating these leads to a bit string of length $n_\text{bits}=n\, m\, \nu$ which describes the entire matrix $\cX$.
With these conventions the phenotype-genotype map is, in fact, bijective and the environment $E$ contains a total of 
\begin{equation}\label{size}
    2^{n_\text{bits}}=2^{nm\nu}
\end{equation}
states.
To orient ourselves let us consider polytopes in dimension $n=5$ with the minimal number, $m=6$, of generators and with each integer $x_a^i$ represented by $\nu=3$ bits (so that, choosing for example $x_\text{min}=-4$, the integer range is $x_a^i\in\{-4,\ldots ,3\}$).
In this case, the environment consists of $2^{90}\simeq 10^{27}$ states, quite a sizeable number and certainly well beyond systematic search.

\subsection{Estimating the Number of Reflexive Polytopes}\label{sec:GLnZ}

We wish to estimate how many states in the total environment correspond to $n$-dimensional reflexive polytopes, given the number of unique reflexive polytopes. 
To do so we estimate the number, $N$, of $\text{GL}(n,\mathbb{Z})$ matrices which transform the unique reflexive polytopes to an equivalent reflexive polytope than remains within the environment, i.e. with vertex coordinates in the range $[x_{\text{min}},x_{\text{min}}+2^{\nu}-1]$. The total number of reflexive polytopes in the environment is then estimated to be (\# unique reflexive polytopes)$\times N$.

We estimate $N$ by the following steps:
\begin{enumerate}
    \item Determine the minimum value of $L$ for which the majority of $\text{GL}(n,\mathbb{Z})$ matrices that transform the unique polytopes to something within the environment have entries in the range $[-L,L]$. To find $L$ we do the following, starting with $L=2$:
    \begin{enumerate}
        \item Generate a random sample of 10000 $\text{GL}(n,\mathbb{Z})$ matrices with entries in the range $[-L,L]$, and either the maximum entry equal to $L$ or the minimum entry equal to $-L$.
        \item Act on the unique reflexive polytopes with each of the 10000 $\text{GL}(n,\mathbb{Z})$ matrices from the previous step and count what number of transformed polytopes exist within the environment, i.e. have vertex coordinates in the range $[x_{\text{min}},x_{\text{min}}+2^{\nu}-1]$. 
        \item If any transformed polytopes remain within environment, increase $L$ by 1 and repeat, otherwise stop.
    \end{enumerate}
    \item Compute the total number, $N_{\text{GL}}$, of $\text{GL}(n,\mathbb{Z})$ matrices with entries in the range $[-L,L]$\footnote{In general, the size of $\text{GL}(n,\mathbb{F}_{p})$, where $\mathbb{F}_{p}$ is a finite field with $p$ elements, is given as $(p^{n}-1)(p^{n}-p)(p^{n}-p^{2})\cdots(p^{n}-p^{n-1})$. Therefore the number of $\text{GL}(n,\mathbb{Z})$ matrices with entries in the range $[-L,L]$ is $((2L+1)^{n}-1)((2L+1)^{n}-(2L+1))((2L+1)^{n}-(2L+1)^{2})…((2L+1)^{n}-(2L+1)^{n-1})$.}:
    \begin{equation}
        N_{\text{GL}} = ((2L+1)^{n}-1)((2L+1)^{n}-(2L+1))\cdots((2L+1)^{n}-(2L+1)^{n-1}).
    \end{equation}
    \item Take the random sample of 10000 $\text{GL}(n,\mathbb{Z})$ matrices with entries in the range $[-L,L]$ from the previous step and determine what fraction $f_{1}\in[0,1]$ of these matrices have determinant $\pm1$, a condition that is necessary to maintain reflexivity.
    \item Generate a sample of 10000 $\text{GL}(n,\mathbb{Z})$ matrices, with entries in the range $[-L,L]$ and determinant $\pm1$, act on the unique reflexive polytopes with each of these matrices and determine what fraction $f_{2}\in[0,1]$ of the transformed polytopes remain within the environment, i.e. have vertex coordinates in the range $[x_{\text{min}},x_{\text{min}}+2^{\nu}-1]$.
    \item Estimate $N$ as 
    \begin{align}
        N = N_{\text{GL}} \times f_{1} \times f_{2}.
    \end{align}
\end{enumerate}

The $N_{\text{GL}}$, $f_{1}$, $f_{2}$ and $N$ values for the two, three and four-dimensional cases are given in Table~\ref{tab:Ntot}.

\begin{table}[h!]
\centering
 \begin{tabular}{|| c c c c c c c c c c||} 
 \hline
 $n$ & $N_{\text{v}}$ & $N_{\text{p}}$ & $x_{\text{min}}$ & $\nu$ & $L$ & $N_{\text{GL}}$ & $f_{1}$ & $f_{2}$ & $N$ \\ [0.5ex] 
 \hline\hline
 $2$ & & & $-3$ & $3$ & $4$ & $5760$ & $0.0433$ & $0.430$ & $107$ \\ 
 $3$ & & & $-7$ & $4$ & $9$ & $3.05\times10^{11}$ & $ 0.000120$ & $0.043$ & $1.58 \times 10^{6}$ \\
 $4$ & 5 & & $-15$ & $5$ & $15$ & $7.03\times10^{23}$ & $4.70\times10^{-8}$ & $1.99\times10^{-5}$ & $6.58\times10^{11}$ \\
 $4$ & & $6$ & $-3$ & $3$ & $4$ & $1.62\times10^{15}$ & $0.000303$ & $0.0349$ & $1.72\times10^{10}$ \\
 $4$ & & $7$ & $-3$ & $3$ & $4$ & $1.62\times10^{15}$ & $0.000303$ & $0.0287$ & $1.41\times10^{10}$ \\
 $4$ & & $8$ & $-3$ & $3$ & $4$ & $1.62\times10^{15}$ & $0.000303$ & $0.0203$ & $1.00\times10^{10}$ \\
 $4$ & & $9$ & $-3$ & $3$ & $4$ & $1.62\times10^{15}$ & $0.000303$ & $0.0144$ & $7.09\times10^{9}$ \\
 $4$ & & $10$ & $-3$ & $3$ & $4$ & $1.62\times10^{15}$ & $0.000303$ & $0.0107$ & $5.27\times10^{9}$\\[1ex] 
 \hline
 \end{tabular}
 \caption{\sf Components of the estimates for the total number of reflexive polytopes that exist in the GA environment. The dimension is given in the first column, the number of vertices and points, if specified, are given in the second and third columns respectively, the minimum coordinate entry is in the fourth column and the maximum number of generators in the fifth column. The range $[-L,L]$ of $\text{GL}(n,\mathbb{Z})$ entries is given in the sixth column, the number of such matrices is given in the seventh column, the fraction of these matrices with determinant $\pm1$ is given in the eighth column, the fraction of $\text{GL}(n,\mathbb{Z})$ matrices with entries in the range $[-L,L]$ and with determinant $\pm1$ which transform the unique reflexive polytopes to polytopes within the environment is given in ninth column and the last column gives the estimate on the total number of reflexive polytopes in the environment.}
\label{tab:Ntot}
\end{table}

\section{Examples}\label{sec:examples}
\begin{example}
    A new five-dimensional reflexive polytope giving rise to a four-dimensional CY hypersurface with $h^{1,1}=1$ is given by the vertex matrix:
    \begin{equation}
        \left(\begin{array}{rrrrrr}
            0 & 4 & -1 & 2 & -1 & -2 \\
            3 & -1 & -4 & 9 & 0 & -3 \\
            2 & 2 & -2 & 2 & -2 & 2 \\
            1 & -3 & 0 & 3 & 0 & -1 \\
            3 & -5 & 1 & -1 & 1 & -1
        \end{array}\right) ~.
    \end{equation}
    This polytope\footnote{In five dimensions, that is for CY four-folds, these topological invariants are not all independent. We have the two relations 
    $h^{2,2} = 44 + 4 h^{1,1} - 2 h^{1,2} + 4 h^{1,3}$ and $\chi = 48 + 6 h^{1,1} - 6 h^{1,2} + 6 h^{1,3}$.
    Mirror symmetry exchanges $h^{1,1}$ and $h^{1,3}$ while leaving $h^{1,2}$ and $h^{2,2}$ fixed.
    } has Hodge numbers $h^{1,2}=0$, $h^{1,3}=111$, $h^{2,2}=492$, and Euler number $\chi=720$.
\end{example}

\begin{example}
    A new five-dimensional reflexive polytope giving rise to a four-dimensional CY hypersurface with $h^{1,1}=2$ is given by the vertex matrix:
    \begin{equation}
        \left(\begin{array}{rrrrrrr}
            0 & -2 & -4 & -1 & 10 & 8 & -2 \\
            4 & 0 & 2 & -1 & -6 & -4 & 0 \\
            -1 & 1 & -3 & 0 & 1 & 3 & 1 \\
            2 & -2 & 0 & 0 & -2 & 0 & 1 \\
            2 & 0 & -2 & 1 & -6 & -2 & 3
        \end{array}\right) ~.
    \end{equation}
    This polytope has Hodge numbers $h^{1,2}=0$, $h^{1,3}=111$, $h^{2,2}=496$, and Euler number $\chi=726$.
\end{example}

\begin{example}
    A new five-dimensional reflexive polytope giving rise to a four-dimensional CY hypersurface whose Euler number $\chi$ is divisible by $24$, $224$, and $504$ is given by the vertex matrix:
    \begin{equation}
        \left(\begin{array}{rrrrrrrr}
            -1 & 0 & 2 & 0 & 2 & -1 & 0 & 1 \\
            0 & -1 & 2 & 0 & 2 & 0 & 0 & 0 \\
            0 & 0 & 0 & -1 & 3 & 0 & 2 & 2 \\
            0 & 0 & 1 & 0 & 0 & 0 & -1 & 0 \\
            -1 & 1 & 0 & 0 & 0 & 0 & 0 & 0 \\
        \end{array}\right) ~.
    \end{equation}
    This polytope has Hodge numbers $h^{1,1}=331$, $h^{1,2}=9$, $h^{1,3}=6$, $h^{2,2}=1374$, and Euler number $\chi=2016$.
\end{example}

\bibliographystyle{elsarticle-num} 
\bibliography{references}

\end{document}